\documentclass[floats,prb,twocolumn,amsmath,superscriptaddress,floatfix]{revtex4}

\usepackage{graphicx}
\usepackage{color}
\usepackage{dcolumn}

\begin{document}

\newcommand{\ket}    [1]{{|#1\rangle}}
\newcommand{\bra}    [1]{{\langle#1|}}
\newcommand{\braket} [2]{{\langle#1|#2\rangle}}
\newcommand{\bracket}[3]{{\langle#1|#2|#3\rangle}}

\def\pw{^{({\rm W})}}
\def\ph{^{({\rm H})}}
\def\k{{\bf k}}
\def\R{{\bf R}}
\def\b{{\bf b}}
\def\q{{\bf q}}
\def\o{{\cal O}}
\def\e{{\cal E}}
\def\v{{\rm v}}
\def\pw{^{({\rm W})}}
\def\ph{^{({\rm H})}}
\def\la{\langle\kern-2.0pt\langle}
\def\ra{\rangle\kern-2.0pt\rangle}
\def\vt{\vert\kern-1.0pt\vert}
\def\D{{D}\ph}
\def\n{{\cal N}}
\def\u{{\cal U}}

\newcommand{\red}{\textcolor{red}}
\newcommand{\blue}{\textcolor{blue}}
\newcommand{\green}{\textcolor{green}}
\newcommand{\cyan}{\textcolor{cyan}}
\newcommand{\magenta}{\textcolor{magenta}}

\title{
Spectral and Fermi surface properties from Wannier interpolation
}

\author{Jonathan R. Yates}
\affiliation{Department of Physics, University of California,
Berkeley, CA 94720}
\affiliation{Materials Science Division, Lawrence Berkeley Laboratory,
Berkeley, CA 94720}

\author{Xinjie Wang}
\affiliation{Department of Physics and Astronomy, Rutgers University,
Piscataway, New Jersey 08854-8019}

\author{David Vanderbilt}
\affiliation{Department of Physics and Astronomy, Rutgers University,
Piscataway, New Jersey 08854-8019}

\author{Ivo Souza}
\affiliation{Department of Physics, University of California,
Berkeley, CA 94720}
\affiliation{Materials Science Division, Lawrence Berkeley Laboratory,
Berkeley, CA 94720}

\date{\today}
\begin{abstract}
We present an efficient first-principles approach for calculating Fermi
surface averages and spectral properties of solids, and use it to compute the
low-field Hall coefficient of several cubic metals and the magnetic circular
dichroism of iron. The first step
is to
perform a conventional first-principles calculation and store the 
low-lying Bloch functions evaluated on a uniform grid of
$k$-points in
the Brillouin zone. We then map those states onto
a set of maximally-localized Wannier functions, and
evaluate the matrix elements
of the Hamiltonian and the other needed operators  between the Wannier
orbitals, thus setting up an  ``exact tight-binding model.''
In this compact representation the $k$-space quantities are evaluated
inexpensively using a generalized Slater-Koster interpolation.
Because of the strong localization of the Wannier orbitals in real space,
the smoothness and accuracy of the $k$-space interpolation increases
rapidly with the number of grid points originally used to construct the 
Wannier functions. 
This allows $k$-space integrals to be performed 
with {\it ab-initio} accuracy
at low cost. In the Wannier representation,
band gradients, effective masses, and other $k$-derivatives needed for
transport and optical coefficients can be evaluated analytically, producing
numerically stable results even at band crossings and near weak
avoided crossings.

\end{abstract}
\vskip 2mm
\pacs{PACS:}

\maketitle

\section{Introduction}
\label{sec:intro}

In electronic structure calculations for solids, the
evaluation of an observable requires integrating
a periodic
function in reciprocal space.\cite{rmm-book} We will
distinguish between
three kinds of properties: those where the integral is
over the Brillouin zone (type-I),
over the Fermi surface
(type-II), and over an energy-difference isosurface (type-III).
In many cases those integrals
take, at $T=0$, the form
\begin{equation}
\label{eq:type1}
I^{(\rm I)}=\sum_n\,\int_{\rm BZ}\frac{d\k}{(2\pi)^3}\,
F_{nn}(\k)\theta(\e_{n\k}-E_{\rm f}),
\end{equation}
\begin{equation}
\label{eq:type2}
I^{(\rm II)}=\sum_n\,\int_{\rm BZ}\frac{d\k}{(2\pi)^3}\,
F_{nn}(\k)\delta(\e_{n\k}-E_{\rm f}),
\end{equation}
and
\begin{equation}
\label{eq:type3}
I^{(\rm III)}(\omega)=\sum_n^{\rm occ}\sum_m^{\rm empty}\,
\int_{\rm BZ}\frac{d\k}{(2\pi)^3}\,
F_{nm}(\k)\delta\big[\hbar\omega-(\e_{m\k}-\e_{n\k})\big].
\end{equation}
Here $E_{\rm f}$ is the Fermi level,
$\e_{n\k}$ are the eigenenergies of the one-electron states, and
$F_{nm}(\k)$ involves matrix elements of
periodic operators which commute with the crystal translations.
Ground-state properties such as the total energy,
and dc response functions such as the Hall conductivity,
are examples of the first and second type of property, respectively.
The third type includes optical absorption in the dipole approximation;
other response and spectral functions can be expressed in a similar
form.\cite{gilat72} Eqs.~(\ref{eq:type1})--(\ref{eq:type3}) are by no means
exhaustive.
While properties such as the electron-phonon interaction [described by matrix
elements of the form $F_{nm}(\k,\k+\q)$ associated with phonon
wavevector $\q$] are not explicitly covered above, the
methods discussed in this paper can be extended to handle such 
cases.\cite{giustino07}

In a practical calculation
the continuous integral is replaced by a summation over a finite number $N$ of
points in the Brillouin zone (BZ),
\begin{equation}
\label{eq:kgrid}
\frac{V_{\rm cell}}{(2\pi)^3}\int_{\rm BZ}\,d\k\rightarrow
\frac{1}{N}\sum_\k\,w(\k) ,
\end{equation}
where $V_{\rm cell}$ is the cell volume and $w(\k)$ are the $k$-point
weights that arise upon restricting the summation to the irreducible wedge
of the BZ. For type-I properties of insulators, the integrand
varies smoothly across the BZ and this summation converges rapidly
with the number of sampled points.\cite{rmm-book}
In metals the BZ integral must be
treated carefully, as the integrand is now discontinuous due to the
partial filling of the energy bands.
Properties of type II and III pose the additional challenge of sampling 
isosurfaces in $k$-space accurately and efficiently.
In all these other cases the rate of convergence is
much slower than for ground state properties of insulators, and a very large
number of $k$-points may be needed.
This sampling problem severely limits the efficiency and accuracy of
{\it ab-initio} calculations for many properties.
Examples of such difficulties abound even in the recent literature,
and include
the magnetocrystalline anisotropy of ferromagnets\cite{tosti03}
and optical
absorption in hot liquid metals,\cite{ping06} 
to name just a few.

In this paper we describe a widely-applicable WF-based scheme
for interpolating both the energy bands $\e_{n\k}$ and the matrix
elements $F_{nm}(\k)$.
The method was used in
Ref.~\onlinecite{xinjie06} to compute
the anomalous Hall conductivity of iron. Where possible, we have
adopted a notation consistent with that work. Ref.~\onlinecite{xinjie06}
dealt with a type-I problem (the quantity being integrated over the
BZ was the Berry curvature of the occupied states), and here
we extend the method to problems of type II and III.

As an example of a type-II problem,
we study the low-field classical Hall coefficient of several cubic
metals.
This and other transport coefficients pose an additional challenge
to existing {\it ab-initio} methodologies: 
how to evaluate the first and possibly
also the second $k$-derivatives of the energy bands at the Fermi level.
Early work\cite{schulz92} employed tight-binding (TB) parameterizations
of {\it ab-initio} bands and the derivatives were calculated by numerical
differentiation  using the linear tetrahedron method.
In other work, an analytic evaluation of the TB band gradients
has been used to achieve
improved numerical stability,\cite{mazin00:_tight_hamil_sr} but the second
derivatives were still computed by finite differences.
Other interpolation
strategies, such as the SKW
scheme\cite{pickett88:_smoot_fourier,uehara00,madsen06}
and spectral differentiation,\cite{chaput:085126}
have also been used. 

All previous interpolation schemes have one feature in common: the only
information retained from the original {\it ab-initio}
calculation is the set of energy eigenvalues on a
grid of $k$-points. Hence the information about the connectivity of the
bands is lost, and the interpolation becomes unreliable or even unstable in
the vicinity of band crossings, avoided crossings, and near-degeneracies.
Moreover, retaining only the eigenenergies strongly restricts the
type of matrix elements $F_{nm}(\k)$, and hence observables,
that can be computed.

A  more powerful interpolation scheme can be obtained
by keeping one more piece of information, namely, the overlap
matrices between the Bloch states at neighboring grid points as in
Eq.~(\ref{eq:overlap}) below. It is perhaps not widely appreciated that the
information about band connectivity is
encoded in those overlap matrices. Indeed, they are the key input for the
WF-construction
method,\cite{marzari97:_maxim_wannier,souza02:_maxim_wannier} and the
connectivity can be recovered from the
Wannier representation of the bandstructure.
Thus, not only do the Wannier-interpolated bands
reproduce the {\it ab-initio} bands with essentially no
loss of accuracy, but their $k$-derivatives
can also be evaluated analytically.
Like the SKW scheme, the present method is based
on Fourier interpolation. Unlike SKW,\cite{uehara00} however,
it produces stable and reliable results even
in the presence of band crossings and avoided crossings.

Remarkably, a knowledge of the overlap matrices of Eq.~(\ref{eq:overlap})
allows for
the interpolation of properties that are not determined
by the energy bands alone, but also depend on the position (or
velocity) matrix elements. 
(More generally, any one-electron operator can be interpolated if, in
addition, its matrix elements
between the WFs are tabulated.)
As an example, we compute the magnetic circular dichroism
of iron, a type-III property.

The paper is organized as follows.
Section~\ref{sec:interpol} contains the methodological aspects of the work.
We start by reviewing the
WF construction methods. We then describe the Wannier-interpolation
strategy for a generic periodic operator.
The interpolation of the velocity operator, as well as of band gradients and
inverse effective masses, is discussed separately.
We conclude Sec.~\ref{sec:interpol} by presenting
an improved broadening scheme for performing the $k$-space integrals.
In Sec.~\ref{sec:hall} we apply the technique to the low-field Hall effect
of several cubic metals, and in Sec.~\ref{sec:mcd} to the
magnetic circular dichroism of bcc Fe.
In Sec.~\ref{sec:conclusion} we provide a brief discussion and conclusion.
The Appendix contains some convergence studies.

\section{Wannier Interpolation}
\label{sec:interpol}

\subsection{Construction of the Wannier functions}
\label{sec:wf_gen}

{\it Ab-initio} calculations provide a certain number
of low-lying Bloch eigenstates $\psi_{n\q}({\bf r})=
e^{i\q\cdot{\bf r}}u_{n\q}({\bf r})$
on a mesh of $k$-points in the BZ, which we take to be uniform.
We will denote points on this ``{\it ab-initio} mesh'' by $\q$, to distinguish
them from arbitrary or interpolation points, which will be denoted by $\k$.

Consider
a type-II (Fermi-surface) problem; two situations may occur. The
first one, which is seen in Pb
for example, occurs when the Fermi level lies within an isolated group of $M$
bands, where by ``isolated'' we mean separated from all higher and lower bands
by a gap throughout the BZ.
In this case it is possible to construct a set of $M$ WFs per unit cell
spanning the Hilbert space of the
isolated Bloch manifold. This can be done using the method of
Marzari and Vanderbilt\cite{marzari97:_maxim_wannier} to obtain so-called
maximally localized Wannier functions 
for that isolated group
of bands.

The second scenario occurs when the
bands of interest
are ``entangled'' with other bands. Then it is still possible, using
the approach of Souza, Marzari, and Vanderbilt,\cite{souza02:_maxim_wannier}
to construct a small number
$M$ of maximally localized WFs which describe those bands exactly.
The number $M$ of WFs per cell
is now to some extent an adjustable parameter. The first
step is to identify the subspace of states of interest. Usually this is done
by selecting the bands inside an energy window spanning from $E_{\rm min}$ to
$E_{\rm max}$. (For type-I and type-III problems $E_{\rm min}$ is normally in
the gap below the lowest valence bands and the position of $E_{\rm max}$
depends on the problem, but is always above $E_{\rm f}$.
For a type-II problem the only requirement is that
$E_{\rm min} <E_{\rm f} < E_{\rm max}$, and in practice the range is
adjusted so that WFs with good localization and symmetry
properties result.) The number
$N_\q$ of states within this window can vary from one $\q$-point to
another, and we require that $M\geq N_\q$ for all $\q$, so that the
space spanned by the WFs (the ``projected space'') can be chosen to contain as
a subspace all the window states. In the method of
Ref.~\onlinecite{souza02:_maxim_wannier} a second (outer) energy window is
used which encloses the previously defined (inner) window.
At each $\q$, the $M$-dimensional projected space is a subspace
of the $\n_\q$-dimensional space of states contained in the
outer window.  For the special case of an isolated group of $M$
bands it is natural to choose $M=N_\q=\n_\q$

Only two pieces of information from the  {\it ab-initio}
calculation are needed as an input to the WF-generation algorithm:
the $\n_\q$ band-energy eigenvalues ${\cal E}_{n\q}$, and the
$\n_\q\times \n_{\q+\b}$ overlap matrices between
the cell-periodic Bloch eigenstates at neighboring points $\q$ and $\q+\b$,
\begin{equation}\label{eq:overlap}
  S_{nm}(\q,\b)=\langle u_{n\q}|u_{m,\q+\b}\rangle.
\end{equation}
The output consists of an $\n_\q\times M$
matrix ${\cal U}(\q)$ for each $\q$. These matrices relate the original set
of $\n_\q$ {\it ab-initio} Bloch eigenstates selected by the outer window to a
new set of $M$ orthonormal Bloch-like states
\begin{equation}
|u_{n\q}\pw\rangle=\sum_{m=1}^{\n_\q}\,|u_{m\q}\rangle
\u_{mn}(\q)
\end{equation}
that vary smoothly with $\q$.
These states are labeled with a
superscript (W) to indicate that the WFs are obtained from them
by a direct Fourier sum
\begin{equation}
\label{eq:wf}
|n\R\rangle=\frac{1}{N_0}\sum_\q\,e^{-i\q\cdot\R}|u_{n\q}\pw\rangle,
\end{equation}
where the sum runs over a grid of $N_0$ $\q$-points and
$|n\R\rangle$ is the $n$-th
Wannier function in the unit cell at $\R$.

Although the explicit construction of the WFs obviously requires a knowledge
of the $|u_{n\q}\rangle$'s, only the
eigenvalues
${\cal E}_{n\q}$  and the overlaps $S(\q,\b)$ are needed to obtain the
$\u(\q)$ matrices. Retaining this minimal information from the {\it ab-initio}
calculation is thus sufficient for many applications, including the
ones presented in this work.

An important object in what follows is the
$M\times M$ Hamiltonian matrix in the projected subspace,
\begin{eqnarray}
\label{eq:hamW}
H_{nm}\pw(\q)&=&\langle u_{n\q}\pw | \hat H(\q) | u_{m\q}\pw
\rangle\nonumber\\
&=&\big[\u^\dagger(\q)H(\q)\u(\q)\big]_{nm},
\end{eqnarray}
where $H_{nm}(\q)={\cal E}_{n\q}\delta_{nm}$ is a diagonal
$\n_\q\times\n_\q$ matrix and
$\hat{H}(\q)=e^{-i\q\cdot\hat{\bf r}}\hat{H}e^{i\q\cdot\hat{\bf r}}$.
We diagonalize $H\pw(\q)$  by finding an $M\times M$ unitary matrix $U(\q)$
such that
\begin{equation}
\label{eq:Htrans}
U^\dagger(\q) H\pw(\q) U(\q) = H\ph(\q),
\end{equation}
where $H\ph_{nm}(\q)={\cal E}\ph_{n\q}\delta_{nm}$. Then ${\cal E}\ph_{n\q}$
will be identical to the original {\it ab-initio} ${\cal E}_{n\q}$ for all
bands inside the inner window.  The corresponding Bloch states
\begin{equation}
\label{eq:twist}
|u_{n\q}\ph\rangle = \sum_m |u_{m\q}\pw\rangle U_{mn}(\q)
\end{equation}
will also coincide with the {\it ab-initio} eigenstates $|u_{n\q}\rangle$
inside the inner window. We shall refer to a quantity with a (W) or (H)
superscript as belonging to the Wannier or Hamiltonian gauge respectively.

\subsection{Wannier interpolation of a periodic operator}
\label{sec:wanninterp}

The problem we pose for ourselves is the following one.  Suppose
we are given a periodic operator
operator $\hat{\o}$,
and we have computed at every $\q$
\begin{equation}
\label{eq:op_abinitio}
\o_{nm}(\q)=\langle u_{n\q}\vert\hat{\o}(\q)\vert u_{m\q}\rangle,
\end{equation}
its matrix elements
between the $\n_\q$ {\it ab-initio} eigenstates in the outer energy window.
How can we interpolate this matrix onto an arbitrary
point $\k$?
We now show that this can be achieved once the
matrices $\u(\q)$ and the eigenvalues ${\cal E}_{n\q}$
($n=1,\ldots,\n_\q$) are known.
Naturally, we can only expect the
interpolation onto a given $\k$ to be meaningful for those matrix elements
$(n,m)$ for which both ${\cal E}_{n\k}$ and ${\cal E}_{m\k}$ fall within the
inner window.

We start by describing in Sec.~\ref{sec:gauge_cov}
the interpolation strategy as it applies to most (``conventional'')
properties. Transport and optical properties merit a separate discussion,
given in Sec.~\ref{sec:non_gauge_cov}.

\subsubsection{
Conventional properties
}
\label{sec:gauge_cov}

By analogy with Eq.~(\ref{eq:hamW}), we define the $M\times M$ matrix
\begin{eqnarray}
\label{eq:opW}
\o_{nm}\pw(\q)&=&\langle u_{n\q}\pw | \hat \o(\q) | u_{m\q}\pw
\rangle\nonumber\\
&=&\big[\u^\dagger(\q)\o(\q)\u(\q)\big]_{nm}.
\end{eqnarray}
Next we find its Fourier sum
\begin{equation}
\label{eq:opWR}
\o_{nm}\pw(\R)=\frac{1}{N_0}\sum_\q\,
e^{-i\q\cdot\R}\o_{nm}\pw(\q).
\end{equation}
This operation is done once and for all for each of the $N_0$ lattice
vectors $\R$ lying in a supercell conjugate to the $\q$-mesh.
(If the sum  is performed using a fast Fourier transform (FFT),
the vectors $\R$ will be disposed in a parallelepipedal supercell.) 
Using Eq.~(\ref{eq:wf}) we recognize in $\o_{nm}\pw(\R)$ the matrix
element of $\hat \o$ between WFs,
\begin{equation}
\label{eq:opWR_b}
\o_{nm}\pw(\R)=\langle n{\bf 0}\vert\hat{\o}\vert m\R\rangle.
\end{equation}

In the above equations, the specification of the lattice vectors $\R$
can be left ambiguous with respect to supercell translations
($\R\rightarrow\R+\R_{\rm sup})$ since $\exp(i\q\cdot\R_{\rm sup})=1$
for all mesh points $\q$, and thus $\o\pw(\R+\R_{\rm sup})=\o\pw(\R)$.
However, we now wish to perform the inverse (slow) Fourier transform
\begin{equation}
\label{eq:op_k}
\o\pw_{nm}(\k)=\sum_\R e^{i\k\cdot\R}\o\pw_{nm}(\R),
\end{equation}
which yields the interpolation of Eq.~(\ref{eq:opW}) onto an
arbitrary point $\k$.
At this point the set of lattice vectors 
must be defined more precisely, since for points $\k$ not
on the $\q$-mesh $\exp(i\k\cdot\R_{\rm sup})\not=1$, and the smoothness of
interpolation will depend on the choice of set.
Using the FFT parallelepipedal supercell, for example, is generally not 
optimal. 
Instead,  
one wants to choose lattice vectors lying inside
the Wigner-Seitz (WS) supercell centered on the 
origin,\cite{souza02:_maxim_wannier,xinjie06} but the details
may vary (e.g., sharing weights of $\R$-vectors lying on the boundary
of the WS supercell, or truncation to a sphere lying within
the WS supercell).  
In practice the
$\vert O\pw_{nm}(\R)\vert$ decay exponentially with $|\R|$, as expected if 
the WFs are exponentially localized, so the results should not be very 
sensitive to this choice, 
when using a sufficiently dense $\q$-mesh.
These remarks are illustrated in the Appendix.  

The final step is to transform the matrix of Eq.~(\ref{eq:op_k})
from the Wannier to the Hamiltonian
gauge. To find the required unitary matrix $U(\k)$
we repeat the above steps for $\hat\o=\hat{H}$ to obtain
$H\pw(\k)$. The matrix $U(\k)$ is then given by
Eq.~(\ref{eq:Htrans}), with the replacement $\q\rightarrow \k$.
Finally,
\begin{equation}
\label{eq:gauge_cov}
\o\ph(\k)=U^\dagger(\k)\o\pw(\k)U(\k),
\end{equation}
where $M\times M$ matrix products are implied on the right-hand side.
This solves the problem posed above at the beginning of
Sec.~\ref{sec:wanninterp}.

Once the WF matrix elements of both the operator of interest and
the Hamiltonian  are tabulated, the interpolation onto an arbitrary $k$-point
requires only inexpensive operations on small $M\times M$ matrices.
When $\hat{\o}=\hat{H}$, the present scheme reduces to Slater-Koster
interpolation, with the maximally localized WFs playing the role of the TB
basis orbitals.\cite{souza02:_maxim_wannier}

Fig.~1 shows the interpolated band structure of bcc Fe along
$\Gamma$--H, using the same WFs and computational
details as in Ref.~\onlinecite{xinjie06}. If one were
to superimpose on this plot the energy bands obtained by performing an
{\it ab-initio} calculation for a large number of
points along the same line in $k$-space,
they would be essentially
indistinguishable from the
interpolated ones. Following
Ref.~\onlinecite{lee05}, we indicate with vertical dashed lines
$k$-points on the $\q$-mesh used for constructing the WFs
(an $8\times 8\times 8$ grid in the full BZ).
It is apparent that the Wannier
interpolation procedure {\it succeeds in resolving details
on a scale much smaller than the spacing between those points.}
In particular, {\it the correct band connectivity is obtained.}
This means that spin-orbit-induced
avoided crossings, for example,
are never mistaken for actual crossings, no matter how weak
the spin-orbit interaction.

\begin{figure}
\includegraphics[width=\columnwidth]{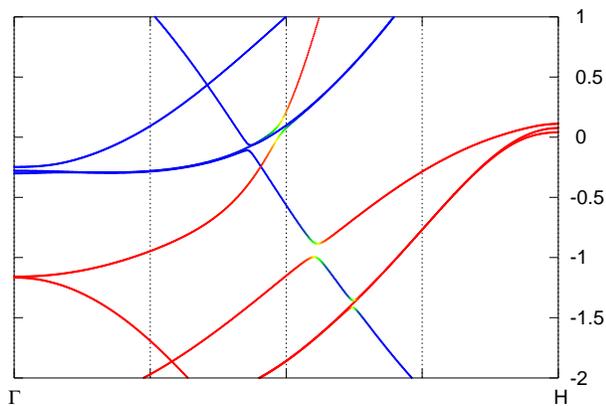}
\caption{\label{fig:fe-bands}
(Color online.)
Wannier-interpolated bands of bcc Fe along $\Gamma$--H. The bands are
color-coded according to the value of $\langle S_z\rangle$: red for spin up
and blue for spin down. The energies are given
in eV and the Fermi
level is at 0~eV. The vertical dashed lines indicate $k$-points on the
{\it ab-initio} mesh used for constructing the WFs.}
\end{figure}

We note in passing that
one could have formulated the problem
at the beginning of Sec.~\ref{sec:wanninterp}
somewhat differently: rather than viewing the matrix elements of $\hat \o$ and
$\hat H$ between the $\n_\q$ {\it ab-initio} Bloch states at each $\q$
as the basic ingredients of the method, we could have assigned that role
to the matrix elements of the two operators between the
WFs. Even if the latter viewpoint is in some ways the more
fundamental one, in practical implementations one often obtains the Wannier
matrix elements (\ref{eq:opWR_b}) via
Eqs.~(\ref{eq:op_abinitio})--(\ref{eq:opWR}).
When doing so, the Wannier orbitals are never explicitly constructed.

\subsubsection{
Transport and optical properties
}
\label{sec:non_gauge_cov}

The treatment of transport and optical properties in crystals is more
subtle. We will restrict our discussion to the electric-dipole approximation, 
where those properties depend on matrix elements of the
velocity operator. The formulation of the previous subsection could in
principle be used to
interpolate the full velocity operator
$\hat v_\alpha=-(i/\hbar)[\hat r_\alpha,\hat H]$
($\alpha=1,2,3$). Its matrix elements, as those of any
other periodic operator, transform between the
Wannier and Hamiltonian gauges according to Eq.~(\ref{eq:gauge_cov}) (such
a matrix will be called ``gauge-covariant''\cite{xinjie06}).
They are given by\cite{blount62}
\begin{equation}
\label{eq:vel}
v_{nm,\alpha}(\k)=\langle \psi_{n\k}|\hat v_\alpha|\psi_{m\k}\rangle=
\frac{1}{\hbar}\,
\Big\langle u_{n\k}\Big|\frac{\partial \hat H(\k)}{\partial k_\alpha}\Big|
u_{m\k}\Big\rangle.
\end{equation}

However,
when describing the dynamics of electrons in crystals it is convenient,
from the points of view of both numerics and physics, to
decompose the velocity operator into two terms.\cite{blount62}
This is achieved
by taking $\partial_\alpha =\partial/\partial k_\alpha$
outside the brackets on the right-hand side of Eq.~(\ref{eq:vel}) and
compensating for the extra terms that appear. After a few manipulations
one obtains
\begin{eqnarray}
\label{eq:vel_b}
v_{nm,\alpha}(\k)=\frac{1}{\hbar}
\frac{\partial{\cal E}_{n\k}}{\partial k_\alpha}\delta_{nm}
-\frac{i}{\hbar}({\cal E}_{m\k}-{\cal E}_{n\k})A_{nm,\alpha}(\k)
\end{eqnarray}
where
\begin{equation}
\label{eq:Awg}
A_{nm,\alpha}(\k)=i\langle u_{n\k}\vert\partial_\alpha u_{m\k}\rangle.
\end{equation}

The first (second) term on the right-hand side of Eq.~(\ref{eq:vel_b})
is diagonal (off-diagonal) in the band index. 
Note that neither
is separately
gauge-covariant. For example, the second one contains $k$-derivatives acting
on the  eigenstates. According to Eq.~(\ref{eq:twist}),
additional terms involving $\partial U(\k)/\partial k_\alpha$ will
therefore appear when transforming between the Wannier and Hamiltonian gauges.
Let us define, for every matrix object~${\cal O}$,
\begin{equation}
\label{eq:utrans}
\overline{\cal O}^{(\rm H)}=
U^\dagger {\cal O}^{(\rm W)}U
\end{equation}
so that, by definition,
$\overline{\cal O}^{(\rm H)}={\cal O}^{(\rm H)}$
only for gauge-covariant objects.
This notation will be used in the next section for expressing
the intraband (diagonal) velocity matrix elements and the
effective mass tensor,
and in Sec.~\ref{sec:mcd} for the interband (off-diagonal)
velocity.

\subsection{Band gradient and Hessian}
\label{sec:grad_hess}

\subsubsection{Notation}
We make use of the first and second $k$-derivatives of the
Hamiltonian matrix,
\begin{equation}
H_{nm,\alpha}=\frac{\partial H_{nm}}{\partial k_{\alpha}},
\end{equation}
\begin{equation}
H_{nm,\alpha\beta}=
  \frac{\partial^2 H_{nm}}{\partial k_{\alpha}\partial k_{\beta}},
\end{equation}
and define $\overline H_{nm,\alpha}$ and $\overline H_{nm,\alpha\beta}$
via Eq.~(\ref{eq:utrans}) as usual.  We also define the first and
second $k$-derivatives of the band energy,
\begin{equation}
\label{eq:band_vel}
\v_{n\k,\alpha}=\frac{1}{\hbar}\frac{\partial {\cal E}_{n\k}}
{\partial k_\alpha},
\end{equation}
\begin{equation}
\label{eq:mass}
\mu_{n\k,\alpha\beta}=\frac{1}{\hbar^2}\frac{\partial^2 {\cal E}_{n\k}}
{\partial k_\alpha\partial k_\beta},
\end{equation}
which have the interpretation of group velocity 
(ignoring Berry-curvature contributions) and inverse effective
mass tensor, respectively.
The strategy for interpolating these quantities is similar to the one
developed in Ref.~\onlinecite{xinjie06} for the Berry
curvature.
We will again make extensive use of the
antihermitian matrix
\begin{equation}
\label{eq:ddef}
\D_{nm,\alpha}\equiv (U^{\dagger}
\partial_{\alpha}U)_{nm}=
\begin{cases}
  \displaystyle
  \frac{\overline H_{nm,\alpha}^{(\rm H)}}{{\cal E}^{(\rm H)}_{m}
  -{\cal E}_{n}^{(\rm H)}}& \text{if $n\not= m$}\\ \\
  0& \text{if $n=m$}
\end{cases}
\end{equation}
defined in that work.

\subsubsection{Non-degenerate bands}

First we consider the band-gradient velocity, Eq.~(\ref{eq:band_vel}).
In the Hamiltonian gauge
$H\ph_{nm}={\cal E}\ph_n\delta_{nm}$, and hence
$H\ph_{nm,\alpha}=\hbar\v\ph_{n\alpha}\delta_{nm}$.
Differentiating Eq.~(\ref{eq:Htrans}) with respect to $k_\alpha$,
\begin{eqnarray}
\label{eq:hder}
H\ph_\alpha
&=&U^\dagger H\pw_\alpha U+
\big\{
  U^\dagger H\pw\partial_\alpha U+{\rm h.c.}
\big\}\nonumber\\
&=&\overline{H}\ph_\alpha+
\big\{
  H\ph D\ph_\alpha+{\rm h.c.}
\big\},
\end{eqnarray}
where each object is an $M\times M$ matrix and
h.c.\ denotes the Hermitian conjugate.
Because of the extra terms in curly brackets we have $H\ph_\alpha\not=
\overline{H}\ph_\alpha$ and thus $H_\alpha$, the first term on the right-hand
side of Eq.~(\ref{eq:vel_b}), is not gauge-covariant.
However, those extra terms do not contribute to the
velocity; being the product of a diagonal matrix ($H\ph$) with an
antihermitian matrix ($D\ph_\alpha$), they only contain off-diagonal elements
which cancel those in $\overline{H}\ph_\alpha$. Thus
\begin{equation}
\label{eq:vel_interpol}
\hbar\v\ph_{n\alpha}=H\ph_{nn,\alpha}=
\overline{H}\ph_{nn,\alpha}=
\Big[U^\dagger H\pw_\alpha U\Big]_{nn}.
\end{equation}

Differentiating this equation yields
the inverse effective mass tensor (\ref{eq:mass})
\begin{eqnarray}
\label{eq:inv-mass1}
\hbar^2\mu\ph_{n,\alpha\beta}&=&
\Big[\partial_{\beta}\overline{H}\ph_{\alpha}\Big]_{nn}\nonumber\\
&=&\Big[U^\dagger H\pw_{\alpha\beta} U\Big]_{nn}+
\Big\{U^\dagger H\pw_{\alpha} \partial_{\beta}U  + {\rm h.c.}\Big\}_{nn}
\nonumber\\
&=& \overline{H}\ph_{nn,\alpha\beta}+
\Big\{\overline{H}\ph_{\alpha}D\ph_{\beta}+{\rm h.c.}\Big\}_{nn}.
\end{eqnarray}
Unlike Eq.~(\ref{eq:hder}), here the matrix
in curly brackets has nonzero diagonal elements which contribute to
$\mu\ph_{n,\alpha\beta}$.

Eqs.~(\ref{eq:vel_interpol})--(\ref{eq:inv-mass1}) are the desired
expressions for the band derivatives, valid away from degeneracies and
inside
the inner energy window.
They involve the $M\times M$ matrices $U(\k)$ calculated
in Sec.~\ref{sec:gauge_cov}, $D\ph_\alpha$ given by
Eq.~(\ref{eq:ddef}), $H\pw_\alpha$, and $H\pw_{\alpha\beta}$. The
last two involve $k$-derivatives of Eq.~(\ref{eq:op_k}) that can
be taken analytically, i.e.,
\begin{equation}
\label{eq:Hder1}
H\pw_{nm,\alpha}(\k)=\sum_\R\, e^{i\k\cdot\R} iR_{\alpha}
\langle n{\bf 0}\vert\hat{H}\vert m\R\rangle
\end{equation}
and
\begin{equation}
\label{eq:Hder2}
H\pw_{nm,\alpha\beta}(\k)=\sum_\R\, e^{i\k\cdot\R} (-R_{\alpha}R_{\beta})
\langle n{\bf 0}\vert\hat{H}\vert m\R\rangle.
\end{equation}

\subsubsection{Discussion}
\label{sec:discussion_method}

In order to interpret the above expressions it is illuminating to
introduce $\vt\phi_n\ra$, the $n$-th
$M$-component column vector of $U$.\cite{xinjie06}
$\vt\phi_n\ra$ is an eigenvector
of $H\pw$, the Hamiltonian operator projected onto the
WF space. We then recognize in
Eq.~(\ref{eq:vel_interpol}) the Hellmann-Feynman result
$\hbar\v\ph_{n\alpha}=\la\phi_n\vt H\pw_\alpha\vt\phi_n\ra$, and in
Eq.~(\ref{eq:inv-mass1}) the expression for the effective mass tensor
in empirical TB theory.\cite{graf95:_elect}
Eq.~(\ref{eq:ddef}) is the standard result from $\k\cdot {\bf p}$
perturbation theory, in terms of the TB states.
It involves the
operator $(1/\hbar)H\pw_\alpha$, which differs from the full
velocity operator
in that the position-operator-dependent terms are 
absent.\cite{graf95:_elect,xinjie06}
We note that all the formulas given so far and in the rest of the paper
remain valid when the {\it ab-initio} Hamiltonian contains
non-local and spin-orbit terms. 

The advantage of this reformulation of ${\bf k}\cdot {\bf p}$
perturbation theory is that it is done strictly in terms of the small number
$M$ of $M$-dimensional states $\vt\phi_n\ra$ at an arbitrary $\k$, and yet it
is exact
within the inner energy window.
In contrast, the formulation in terms of the original
{\it ab-initio} states on the $\q$-grid
is considerably more expensive and usually entails a truncation
error.\cite{pickardkp}

\subsubsection{Degeneracies}
\label{sec:degen}

While meaningful band derivatives can be defined via degenerate
$\k\cdot {\bf p}$ perturbation theory even at degeneracy points in
the BZ,\cite{boykin95:_incor} this is not possible when the only
information available about the bandstructure is a list of eigenenergies
on a predetermined coarse $k$-point grid.  In that case, the
information about the band connectivity is lost, and
finite-difference estimates of the derivatives become ill-defined
at points of degeneracy, which have to be carefully
avoided.\cite{gilat72}

In the present formulation
the $k$-gradients of the degenerate states are the eigenvalues of
the submatrix
 \begin{equation}
\label{eq:degen1}
\Big[\overline{H}\ph_{\alpha}\Big]_{\mu\nu}=
\la\phi_\mu\vt H\pw_\alpha\vt\phi_\nu\ra,
\end{equation}
where the indices $\mu$ and $\nu$ run over the degenerate states only.
We update the $M\times M$ matrix $U$ by replacing the
columns corresponding to those states with the rotated states that
diagonalize $H\pw_\alpha$.
The Hessian matrix can then be obtained from
Eq.~(\ref{eq:inv-mass1}) using the updated $U$ and a modified form of
Eq.~(\ref{eq:ddef}),

\begin{equation}\label{eq:wan-pert2}
D\ph_{nm,\alpha}
=
\begin{cases}
  \displaystyle
  \frac{\overline H_{nm,\alpha}^{(\rm H)}}{{\cal E}^{(\rm H)}_{m}
  -{\cal E}_{n}^{(\rm H)}}& \text{if ${\cal E}^{(\rm H)}_{n}\not= {\cal E}^{(\rm H)}_{m}$}\\ \\
  0& \text{if ${\cal E}^{(\rm H)}_{n}= {\cal E}^{(\rm H)}_{m}$} .
\end{cases}
\end{equation}

In cases of degeneracies (at band edges, for example) where some
of the eigenvalues of the matrix (\ref{eq:degen1}) are equal, a
first-order treatment is inadequate and the correct rotation between
eigenstates needed to compute the Hessian is found by going to second order
in degenerate perturbation theory. This amounts to
diagonalizing the submatrix obtained from the
right-hand side of Eq.~(\ref{eq:inv-mass1}) by replacing the subscripts
$nn$ therein by $\mu\nu$, and
letting $\mu$ and $\nu$ run over the
the first-order-degenerate bands. The desired Hessian
matrix elements are the eigenvalues of that
submatrix.

In our calculations we employ the first-order form of
degenerate perturbation theory when two or more energy eigenvalues lie within
10$^{-4}$~eV of
each other and we subsequently use the second-order form if in addition two
or more band gradients differ by less than 0.1~eV-\AA.

\subsubsection{Validation}
\label{sec:examples}

\begin{figure}
\includegraphics[width=\columnwidth]{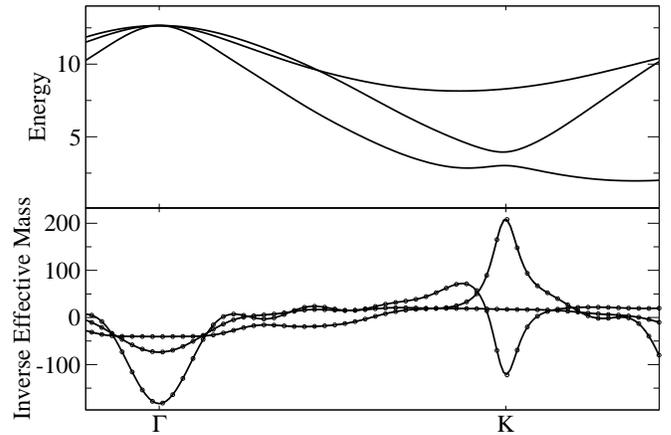}
\caption{\label{fig:lead-curve}
Upper panel: Dispersion of the three $p$-like energy bands of
Pb along the $\Gamma$--K direction, obtained by interpolating a 
non-relativistic {\it ab-initio} calculation. Lower panel: Inverse effective 
masses of those bands along the same direction, calculated in two ways:
from the interpolated eigenenergies on a regular mesh of points using a spline
fit (circles), and from perturbation theory in the Wannier representation
(solid lines).}
\end{figure}
As an illustration, we used the above
formulas to calculate the inverse effective mass of the
three $p$-like valence bands of Pb along the $\Gamma$--K direction in
$k$-space. (We have chosen this example because it displays a threefold
band-edge degeneracy at $\Gamma$ and a band crossing between $\Gamma$ and K.)

In all the calculations in this work, lattice constants are taken from
Ref.~\onlinecite{papaconstantopoulos86:_handb_band_struc_elemen_solid}. The
Bloch states are obtained with the {\tt PWSCF} code\cite{pwscf} using
density-functional theory in the  local-density approximation, together with
the planewave-pseudopotential 
formalism.\cite{rmm-book}
Norm-conserving pseudopotentials are employed, and spin-orbit
effects are included in Sec.~\ref{sec:mcd} and Fig.~\ref{fig:fe-bands} only. 
The WFs are generated using the 
{\tt WANNIER90} code.\cite{wannier90}

In Pb the lowest four valence bands crossing the Fermi level are separated 
everywhere in the
Brillouin zone from  higher bands, so that the original prescription of
Marzari and Vanderbilt\cite{marzari97:_maxim_wannier}
can be used to generate the corresponding maximally localized WFs. 
They are atom-centered and have $sp^3$ character, with the axis of each
orbital pointing towards a nearest neighbor.

The inverse effective masses are shown as solid lines in the lower panel of
Fig.~\ref{fig:lead-curve}. For comparison, we also
plot as circles in the lower panel the values obtained by
fitting a spline function to the energy eigenvalues on
a dense grid of $k$-points along the path.
We remark that whereas in the analytic method band crossings are handled
automatically, in order to obtain a smooth spline fit it was necessary
to manually reorder the eigenvalues close to the band
crossing. Once that is done, the agreement between the two is essentially
perfect.

\subsection{Adaptive broadening scheme for $k$-space integration}
\label{sec:broadening}

We conclude this section by discussing the evaluation of type-I, II, and III
integrals, Eqs.~(\ref{eq:type1}--\ref{eq:type3}). In order to
accelerate the convergence of type-I integrals with respect to the number of
sampling points in Eq.~(\ref{eq:kgrid}),
a broadening scheme can be used.\cite{fu83,methfessel89}
This amounts to replacing
the step function in Eq.~(\ref{eq:type1}) with a Fermi-Dirac-like smearing
function. In the case of type-I integrals, smearing is most important when
relatively few sampling points are used,
as tends to be the case whenever the evaluation of the integrand
is expensive.\cite{methfessel89}
If, however, the integrand is inexpensive,
as is the case when using Wannier interpolation,
then it is possible to converge the BZ integral
without resorting to smearing.
For example, no smearing was used in
Ref.~\onlinecite{xinjie06} for integrating the Berry
curvature over the occupied states of bcc Fe.

Smearing plays a more fundamental role in integrals of types
II and III: when replacing the BZ integral in
Eqs.~(\ref{eq:type2}-\ref{eq:type3})
by a grid summation, the $\delta$-functions must be
replaced by normalized functions with non-zero width, such as Gaussians.
For example, in Eq.~(\ref{eq:type2}) one would replace
$\delta(E_{\rm f}-\e_{n\k})$ by
\begin{equation}
\label{eq:fermi_gauss}
g_{n\k}(E_{\rm f})=\frac{1}{\sqrt{2\pi}W}\exp
\left(\frac{-(E_{\rm f}-\e_{n\k})^2}{2W^2}\right).
\end{equation}
Ideally the Gaussian width should be,
for a given grid spacing $\Delta k$,
comparable with the level spacing $\Delta \e_{n\k}$.
The level spacing is however difficult to estimate, and the common
practice is to set $W$ to a constant for all bands and $k$-points.
As a result, FS
sheets arising from steep and flat bands are not described
consistently. This is a serious disadvantage
of broadening schemes with respect to the linear tetrahedron method,
as discussed in Ref.~\onlinecite{blochl94}.

This drawback is easy to remedy within the Wannier-interpolation method, since
the band derivatives are readily available (Sec.~\ref{sec:grad_hess}), and can
be used to estimate the level spacing. The simple estimate
$\Delta\e_{n\k}\sim |\partial\e_{n\k}/\partial \k|\Delta k$
suggests using a state-depending broadening width
\begin{equation}
W_{n\k}=a\left|\frac{\partial \e_{n\k}}{\partial\k}\right|\Delta k
\end{equation}
for type-I and type-II integrals ($a$ is a dimensionless constant of the
order of unity), and
\begin{equation}
W_{nm,\k}=a
\left|
  \frac{\partial \e_{m\k}}{\partial\k}-\frac{\partial \e_{n\k}}{\partial\k}
\right|\Delta k
\end{equation}
for type-III integrals. With this prescription $W$ is no
longer an independent adjustable parameter from  $\Delta k$, guaranteeing that
the $\Delta k\rightarrow 0$ and $W\rightarrow 0$ limits are approached
consistently. Several smearing functions beyond a simple Gaussian have
been proposed\cite{marzari97:_ensem_densit_funct_theor_ab,methfessel89} and can be used straight forwardly with
the adaptive smearing scheme. For all of the calculations presented in
this work we use the first-order Hermite polynomial scheme introduced by
Methfessel and Paxton.\cite{methfessel89} 

\begin{figure}
\includegraphics[width=\columnwidth]{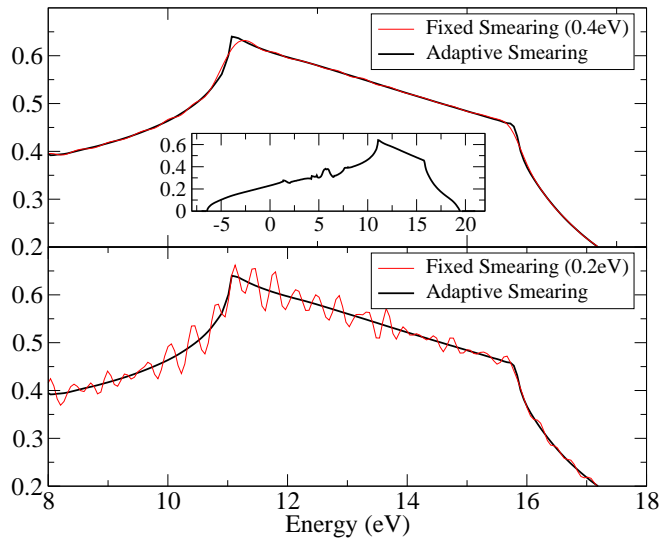}
\caption{\label{fig:smearing}
(Color online.)
Density of states of bulk diamond calculated in the range 8-18eV using the conventional
Gaussian broadening approach (light thin lines) with a fixed width of
0.4~eV in the upper panel and 0.2~eV in the lower panel, versus the adaptive
broadening approach (dark thick lines). The inset shows the density of
states in the full valence band range computed with the adaptive
broadening approach.}
\end{figure}

The above first-order adaptive smearing should
be reliable whenever the level-spacing is gradient-dominated.
In practice we find that it works rather well
even 
near critical points.This is
illustrated in Fig.~\ref{fig:smearing}, where we show the density of states
of diamond calculated using both the adaptive and conventional
(fixed width) smearing, with a $50\times50\times50$ interpolation mesh.
When using a fixed width of $W=0.4$~eV, the sharp
van-Hove features are not well described. Reducing it to 0.2~eV
improves the situation for some of them, but introduces
spurious oscillations whenever the level-spacing becomes larger than $W$.
With the adaptive scheme such oscillations do not occur for a sensible choice
of $a$,
and the sharp
features are well-described. We have used $a=1.0$, but find the results to be
quite robust for $0.8<a<1.3$.

\section{Low-field Hall coefficient of cubic metals}
\label{sec:hall}

As a first benchmark application
we compute the  ``classical'' low-field Hall coefficient
of cubic metals, which is given by
\begin{equation}
\label{eq:RH}
R_{\rm H} = \frac{\sigma_{xy,z}}{\sigma_{xx}^2},
\end{equation}
\begin{equation}
\label{eq:sigma_xx}
\sigma_{xx} =q_e^2\sum_n\int \frac{d\k}{(2\pi)^3}\,\tau_{n\k}
\v_{n\k,x}^2\left(-\frac{\partial f}{\partial {\cal E}}\right),
\end{equation}
and
\begin{eqnarray}
\label{eq:sigma_xy}
\sigma_{xy,z}=&q_e^3&\sum_n\int \frac{d\k}{(2\pi)^3}\,
\tau_{n\k}^2
\left(-\frac{\partial f}{\partial {\cal E}}\right)\times\nonumber\\
&\times&
\left(
  \v_{n\k,x}^2 \mu_{n\k,yy}-\v_{n\k,x}\v_{n\k,y}\mu_{n\k,xy}
\right).
\end{eqnarray}
(For the systems studied in this Section, which are non-ferromagnetic
and do not include the spin-orbit interaction, the sum over spin-degenerate 
bands will give rise to factors of two, which are not written explicitly.)
$\sigma_{xx}$ is the longitudinal conductivity, 
and and $\sigma_{xy,z}=\partial\sigma_{xy}/\partial B_z$, where
$\sigma_{xy}$ is the antisymmetric (Hall) conductivity.
The above expressions are 
obtained from a Bloch-Boltzmann description of transport;
for a detailed derivation, see Ref.~\onlinecite{hurd72}. We note that
Eq.~(\ref{eq:sigma_xy}) is written in a form which is specific to cubic
metals.
The quantities $\v_{n\k,\alpha}$ and $\mu_{n\k,\alpha\beta}$ are given by
Eqs.~(\ref{eq:band_vel}) and (\ref{eq:mass}),
$f({\cal E})$ is the Fermi-Dirac distribution function, 
and $q_e<0$ is the electron charge.
At low temperatures $(-\partial f/\partial{\cal E})$
tends to $\delta(\e-E_{\rm f})$, and both
$\sigma_{xx}$ and $\sigma_{xy,z}$ become FS integrals of the
form of Eq.~(\ref{eq:type2}).
In the constant relaxation-time approximation
$\tau_{n\k}$ drops out from
Eq.~(\ref{eq:RH}) so that $R_{\rm H}$ is completely specified by the first and
second band derivatives at $E_{\rm f}$.

Calculations were done for Li, Al, Cu and Pd. Unlike Pb, in these metals the 
set of bands crossing the Fermi level is not isolated. Therefore, in order to 
generate maximally localized WFs we first used the disentanglement 
procedure summarized in Sec.~\ref{sec:wf_gen} to obtain an optimal
projected space. The number of bands contained therein
must be at least equal to the number of bands crossing the Fermi level. 
However, it is often desirable to extract a somewhat larger projected space if 
this produces a more symmetric set of Wannier functions. 

For lithium we obtained four atom-centered WFs per primitive cell with
$sp^3$ character. For aluminum we extracted a nine-dimensional projected 
subspace. The resulting WFs are atom-centered,
but have no distinct symmetry characteristics. For Cu and Pd we used
seven WFs: five with $d$ character on atom centers, and two with $s$
character located at the tetrahedral interstitial sites. These have been
previously described for Cu in Ref.~\onlinecite{souza02:_maxim_wannier}.

The computational details are the same as in 
Section \ref{sec:examples}.
We obtain the
self-consistent ground state using a 16$\times$16$\times$16
Monkhorst-Pack mesh of $k$-points and a fictitious Fermi
smearing\cite{methfessel89} of 0.02\,Ry for the Brillouin-zone
integration. We use the local density approximation; for the materials
studied we find the results are not significantly altered by
using a GGA such as PBE.\cite{perdew1}
\begin{table}
\caption{
\label{table:hall}
Hall coefficient $R_{\rm H}$, in units of $10^{-11}$m$^3$C$^{-1}$.
References to the experimental data can be found in Ref.~\onlinecite{hurd72}.
}
\begin{ruledtabular}
\begin{tabular}{lcccc}
& This work & Ref.~\onlinecite{schulz92} & Ref.~\onlinecite{uehara00} & Experiment \\
\hline
Li & $-$12.7  & $-$12.8 & $-$12.4 & $-$15.5 \\
Al & $-$2.5   & $-$1.7  & $-$3.4  & $-$3.43 \\
Cu &  $-$4.9 & $-$5.2  & $-$4.9 & $-$5.17  \\
Pd & $-$11.9   & $-$6.0  & $-$17 & $-$7.6   \\
\end{tabular}
\end{ruledtabular}
\end{table}
To compute the Hall coefficient we use an {\it
  ab-initio} grid of 12$\times$12$\times$12 {\bf q}-points and obtain the 
required quantities on a uniform mesh of 60$\times$60$\times$60 $\k$-points. 
We implement an adaptive mesh refinement scheme in which we identify those 
points of the $k$-space mesh at which at least one band lies with 0.5eV of the 
Fermi energy and obtain
the required quantities on a $7\times 7\times 7$ submesh spanning the 
original cell associated with this mesh point. We find these parameters give
converged values of the Hall coefficient for the four metals
studied. This is particularly reassuring in the case of Pd, where previous
techniques encountered difficulties because of the occurrence of bands
crossings near the Fermi surface.\cite{uehara00}

The results for the Hall coefficient $R_{\rm H}$ are compiled in 
Table~\ref{table:hall}. For Li, Al, and Cu we find excellent agreement with
previous calculations based on empirical TB fitting to {\it ab-initio}
bands,\cite{schulz92} as well as {\it ab-initio} calculation combined with
SKW interpolation.\cite{uehara00} The case of Pd is more delicate as
$R_{\rm H}$ depends critically on the details of the {\it ab-initio}
calculation. For example a shift upwards (downwards) in the Fermi
level of 0.002Ry causes a change of $-$3\ (+2)\ $\times10^{-11}$m$^3$C$^{-1}$.
In view of this we refine the position
of the Fermi level in a final non-self-consistent step by interpolating
the band energies and gradients onto a $60\times 60\times 60$ $\k$-mesh and 
using the adaptive broadening scheme. Our converged value of $R_{\rm H}$ is
intermediate between the two previously computed values, and shows
a relatively large discrepancy with experiment. Previous authors have
suggested\cite{schulz92,Beaulac82} that it maybe necessary to go
beyond the constant relaxation time approximation to give a good
description of the transport properties of Pd.

\section{Magnetic circular dichroism}
\label{sec:mcd}

Magneto-optical effects in ferromagnets result from 
a combination of exchange splitting and spin-orbit coupling 
(SOC).\cite{ebert96,antonov-book} The former 
breaks time-reversal (TR) in the spin channel, and the latter transmits the
TR-breaking to the orbital motion of the electrons, endowing the 
optical conductivity tensor with an antisymmetric component. 
The simplest such effect to evaluate is magnetic circular dichroism (MCD), the 
difference in absorption between left- and right-circularly-polarized light, 
and we have chosen it for illustrative purposes. It is given by the imaginary 
part of the antisymmetric conductivity, 
$\sigma_{\rm A,\alpha\beta}^{(2)}(\omega)=-\sigma_{\rm A,\beta\alpha}^{(2)}(\omega)$.

\subsection{Evaluation of the Kubo formula}
\label{sec:mo_kubo}

{\it Ab-initio} calculations of magneto-optical effects demand high 
accuracy and dense $k$-space sampling. 
The spin-orbit interaction is typically a small perturbation on top 
of the much larger exchange splitting, and the
modifications that it produces on the electronic structure (both in the
energy bands and in the matrix elements) are subtly and strongly
dependent on $k$-point and band index.

The conductivity $\sigma_{\rm A,\alpha\beta}^{(2)}(\omega)$ is
evaluated from the Kubo formula of linear-response 
theory in the electric-dipole approximation.\cite{antonov-book} 
The needed ingredients are the energy eigenvalues 
of the states involved in the optical transitions and the transition matrix 
elements. We will evaluate the interband contribution to 
the magneto-optical absorption using Eq.~(\ref{eq:vel_b}) for the 
electric-dipole transition matrix elements,
where it is now understood that all Bloch functions $|u_{n\k}\rangle$
are spinors determined from a Hamiltonian that includes the spin-orbit
interaction. One finds
\begin{eqnarray}
\label{eq:sigma_mcd}
\sigma^{(2)}_{{\rm A},\alpha\beta}(\omega)=
&-&\frac{\pi e^2\omega}{\hbar}
\sum_n^{\rm occ}\sum_m^{\rm empty}\int\frac{d\k}{(2\pi)^3}\,
{\rm Im}\big(A_{nm,\alpha} A_{mn,\beta}\big)\nonumber\\
&\times&\left[\delta(\omega-\omega_{mn})-\delta(\omega+\omega_{mn}) \right],
\end{eqnarray}
where $\hbar\omega_{mn}=\e_m-\e_n$.

Eq.~(\ref{eq:sigma_mcd}) is a type-III integral of the form of
Eq.~(\ref{eq:type3}). When evaluating it by Wannier interpolation it must be
kept in mind that the Wannier-derived bands
reproduce the {\it ab-initio} ones only inside
the inner energy window, and therefore its range must be adjusted according to
the maximum desired absorption frequency.
The matrix elements $A_{nm,\alpha}$ are to be evaluated in the
Hamiltonian gauge, and the interpolation of $A_{nm,\alpha}\ph$ is based on the
two relations\cite{xinjie06}
\begin{equation}
A_\alpha\ph=\overline{A}_\alpha\ph+i\D_\alpha
\label{eq:Ata}
\end{equation}
and
\begin{equation}
A_{nm,\alpha}\pw(\k)=\sum_\R e^{i\k\cdot\R}\;
   \langle{\bf 0}n|\hat{r}_\alpha|\R m\rangle\,,
\label{eq:ww}
\end{equation}
where $\D_\alpha$ is given by Eq.~(\ref{eq:ddef}) and
$\overline{A}_\alpha\ph$ and $A_\alpha\pw$ are related by
Eq.~(\ref{eq:utrans}).
Inserting  Eq.~(\ref{eq:Ata}) into
${\rm Im}(\ldots)$ in Eq.~(\ref{eq:sigma_mcd}), we find
\begin{eqnarray}
{\rm Im}\big( A_{nm,\alpha}\ph A_{mn,\beta}\ph\big)&=&
{\rm Im}\left( \overline A_{nm,\alpha}\ph \overline A_{mn,\beta}\ph\right)
\nonumber\\
&+&{\rm Re}\left(\overline A_{nm,\alpha}\ph D_{mn,\beta}\ph+
D_{nm,\alpha}\ph\overline{A}_{mn,\beta}\ph\right)\nonumber\\
&-&{\rm Im}\left(D_{nm,\alpha}\ph D_{mn,\beta}\ph\right).
\label{eq:AA}
\end{eqnarray}
The contributions to $\sigma^{(2)}_{{\rm A},\alpha\beta}(\omega)$ from the
three terms on the right-hand side will be denoted as
$\overline{A}$--$\overline{A}$, $D$--$\overline{A}$, and $D$--$D$,
respectively.

\subsection{Results for bcc Fe}
\label{sec:results_magneto_optical}

\begin{figure}
\includegraphics[width=\columnwidth]{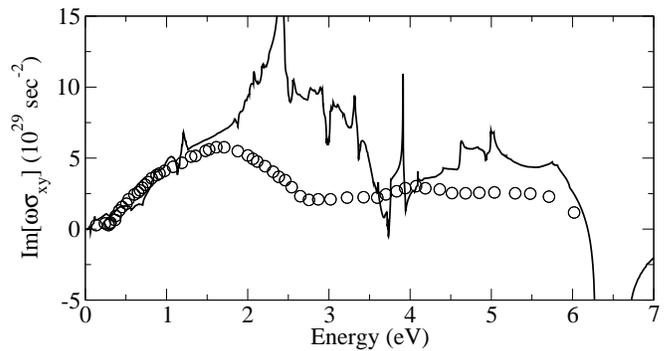}
\caption{\label{fig:mcd}Magnetic circular dichroism spectrum of bcc iron.
The calculated spectrum (solid lines) is compared with the experimental
spectrum from Ref.~\onlinecite{krinchik67} as reproduced in 
Ref.~\onlinecite{yao04} (open circles).}
\end{figure}

\begin{figure}
\includegraphics[width=\columnwidth]{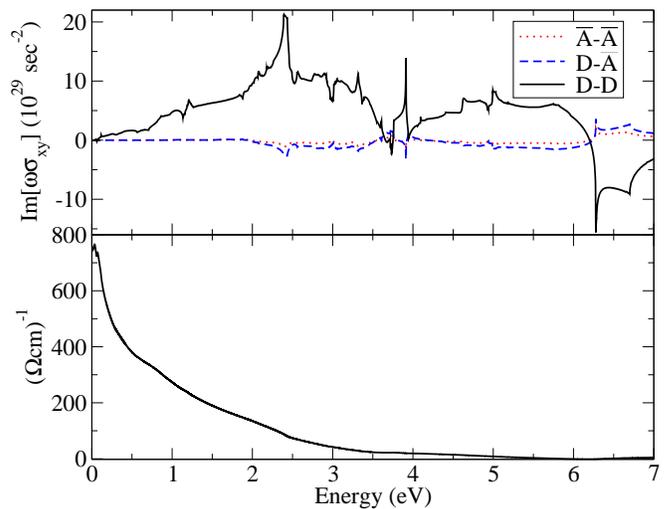}
\caption{\label{fig:mcd_contrib}
(Color online.)
Upper panel: decomposition of the magnetic circular dichroism spectrum
into the three terms defined by the Wannier-interpolation procedure.
Lower panel: cumulative anomalous Hall conductivity (AHC) versus $\omega$, 
defined as the contribution to the AHC 
from frequencies higher than $\omega$ in Eq.~(\ref{eq:kk}).
}
\end{figure}

Unlike the calculations presented earlier in the paper, to calculate the MCD
spectrum we have used relativistic pseudopotentials which explicitly include
spin-orbit effects.\cite{corso05} The computational  details
for the {\it ab-initio} calculation, WF-generation, and treatment of
the spin-orbit interaction are the same as in Ref.~\onlinecite{xinjie06}.
The integral in Eq.~(\ref{eq:sigma_mcd}) was evaluated on a uniform 
$\k$-mesh containing
$125\times 125\times 125$ points using the adaptive broadening
scheme, which we find to be essential for resolving
the fine details in the MCD spectrum.
The spin magnetization is along the $z$-axis, so that $\sigma_{{\rm
    A},xy}^{(2)}(\omega)$ is the only independent non-zero component.

It is conventional to plot the MCD spectrum as
$\omega\sigma_{{\rm A},xy}^{(2)}(\omega)$,
and adopt this convention in Fig.~\ref{fig:mcd}. Our calculated
spectrum for bcc Fe is in excellent agreement with the one computed in
Ref.~\onlinecite{yao04} using a completely different electronic structure 
method. (Previous calculations of 
magneto-optical effects have most commonly used all-electron methods.)
As remarked in Ref.~\onlinecite{ebert96},
this level of agreement between two different calculations
is non-trivial when it comes to the spin-orbit-induced
$\sigma_{{\rm A},xy}^{(2)}(\omega)$. It provides a strong validation
of the Wannier-interpolation scheme combined with the
pseudopotential-planewave method as a viable way of computing
magneto-optical effects. 

The upper panel of
Fig.~\ref{fig:mcd_contrib} shows the decomposition of the calculated MCD
spectrum into the three contributions
($\overline{A}$--$\overline{A}$, $D$--$\overline{A}$, and $D$--$D$)
defined by the Wannier-interpolation procedure, as discussed following
Eq.~(\ref{eq:AA}).
It is clear that the $D$--$D$ contribution tends to dominate the
spectrum in the frequency range from 0 to 7~eV, especially at the
lowest frequencies.  For frequencies above 7~eV (not shown), the
$\overline{A}$--$\overline{A}$ and $D$--$\overline{A}$ terms
become significant.

The interband MCD spectrum 
$\sigma_{{\rm A},xy}^{(2)}(\omega)$ is related to the
Karplus-Luttinger anomalous Hall conductivity\cite{xinjie06} (AHC) 
$\sigma_{{\rm A},xy}^{(1)}(0)$ by the Kramers-Kr\"onig relation  
\begin{equation}
\label{eq:kk}
\sigma_{{\rm A},xy}^{(1)}(0)=\frac{2}{\pi}\int_0^\infty\,
\frac{1}{\omega}\sigma_{{\rm A},xy}^{(2)}(\omega)\,d\omega.
\end{equation}
In the lower panel of Fig.~\ref{fig:mcd_contrib} we show
the cumulative AHC versus $\omega$, defined as the contribution to the AHC 
from frequencies higher than $\omega$ in Eq.~(\ref{eq:kk}). In practice 
we use as the upper
frequency limit in Eq.~(\ref{eq:kk}) the difference from
the Fermi energy to the top of the inner energy window (18~eV). It is clear
that the AHC is completely dominated by the low-frequency contributions
below $\sim5.5$~eV.

It can be shown that applying the transformation (\ref{eq:kk}) separately
to the $D$--$D$ term of the MCD spectrum yields the $D$--$D$ term of the 
AHC, as defined in Ref.~\onlinecite{xinjie06}. This explains the intriguing
result that more than 99\% of the
anomalous Hall conductivity can be recovered from the $D$--$D$
term alone.\cite{xinjie06} This is a consequence of
(i) the low-frequency 
part of the spectrum being weighted more in the integral as a result of the
$1/\omega$ factor in the integrand, and 
(ii) the $D$--$D$ term overwhelming the other two at 
very low frequencies.

\section{Conclusions}
\label{sec:conclusion}

We have presented a Wannier-interpolation scheme to 
compute efficiently and accurately Fermi-surface and
spectral properties from first principles. 
As an example of the former we computed the low-field Hall conductivity for
several cubic metals. As an example of the latter we calculated the
magnetic circular dichroism spectrum of bcc Fe.

The scheme naturally resolves a number of difficulties which have
plagued existing interpolation schemes. Firstly, by preserving the information
about band connectivity, band crossings and avoided crossings are treated 
correctly. In addition, the evaluation of the velocity 
matrix elements needed to compute both the Hall coefficient and the MCD
spectrum can be done analytically in the Wannier
representation. Furthermore, the scheme 
does not become any more complex upon inclusion of the spin-orbit interaction 
in the Hamiltonian. In particular, there are no additional contributions
to the velocity matrix elements; all the spin-orbit-related corrections are
contained in the spinor WFs. Also, the Wannier-interpolation scheme is 
decoupled from the particular choice of basis set used for performing the 
original
{\it ab-initio} calculation, nor does it depend on the  specific level of 
single-particle theory.
As such, the calculation of a given property can be implemented in a universal
way inside the Wannier module, which can then be interfaced with any desired 
electronic structure code.

The appeal of the present approach is that it combines the
simplicity of a tight-binding-like scheme with the power and accuracy of
{\it ab-initio} methods. Most importantly, {\it it allows operators other than 
the Hamiltonian to be interpolated in the same manner as the 
Slater-Koster interpolation of energy bands.} As such, it can be applied to
a wide variety of problems in condensed matter physics. It should be 
particularly useful for studying metallic systems. A number of properties of 
metals remain extremely challenging to compute from first-principles, as a 
result of difficulties in sampling the Fermi surface with sufficient accuracy.
Wannier interpolation provides an elegant and powerful framework for 
investigating such problems with {\it ab-initio} techniques.

\section{Acknowledgments}

This work was supported by the Laboratory Directed Research and
Development Program of Lawrence
Berkeley National Laboratory under the Department of Energy Contract
No.~DE-AC02-05CH11231, and by NSF Grant 0549198.

\appendix*

\section{Convergence properties of the interpolation scheme}
\label{app:conv}
%
\begin{figure}
\includegraphics[width=\columnwidth]{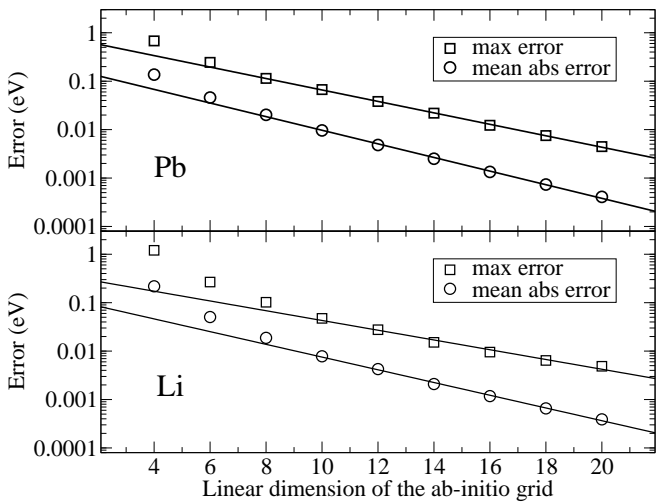}
\caption{\label{fig:band-conv}
Convergence of the Wannier interpolated band energies as a function of
the linear dimensions $N_0^{1/3}$ of the {\it ab-initio} $\q$-point grid. 
We plot the maximum error
(squares) and mean absolute error
(circles), where the error is the difference between the Wannier
interpolated band energy and the value obtained from a full non-self
consistent diagonalization of the planewave Hamiltonian. The lines are
linear fits to the points with $N_0^{1/3}>8$.}
\end{figure}
\begin{figure}
\includegraphics[width=\columnwidth]{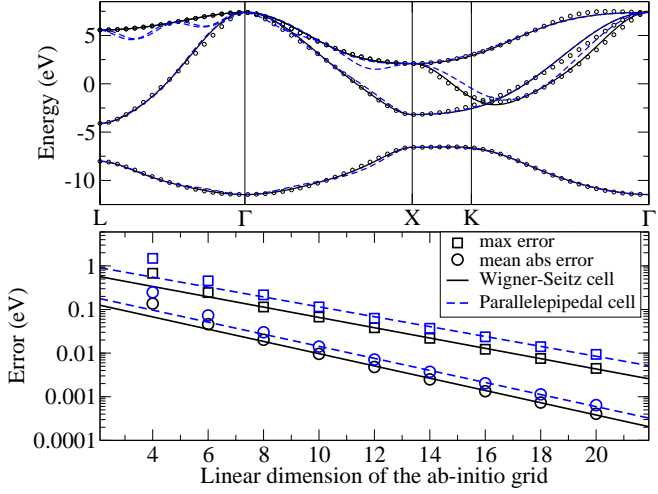}
\caption{\label{fig:wigner_vs_par}
(Color online.) Comparison of interpolated band-energies for Pb
  obtained using sets of lattice vectors defined within Wigner-Seitz
  (WS) and parallelepipedal (P) supercells. Upper figure: Energy bands
  interpolated using a $4\times4\times4$ $\q$-point grid (WS
  cell -- solid lines, P cell -- dashed lines). Full {\it ab-initio} results are
  shown as open circles. Lower figure: Convergence of the Wannier-interpolated
  band energies for the two supercells as a function of the linear dimensions
  $N_0^{1/3}$ of the $\q$-point grid. Details as in Fig.~\ref{fig:band-conv}.}
\end{figure}

For a given operator $\hat{\cal O}$ the
agreement inside the inner energy window between ${\cal O}_{nm}\ph(\k)$ 
obtained by Wannier interpolation 
and ${\cal O}_{nm}(\k)$ calculated using a full first-principles calculation is
determined by $N_0$, the number of points in the $\q$-grid. 
The resulting WFs are 
periodic over the the conjugate real-space supercell spanning
$N_0$ unit cells.
For any finite $N_0$ there is some overlap between a WF and its 
neighboring periodic images, which affects the matrix ${\cal O}(\R)$.
It is generally accepted that WFs decay
exponentially;  numerical studies have confirmed this for several materials,\cite{PhysRevLett.86.5341}
and recently there has been a claim of a formal proof for multiband
time-reversal-invariant insulators.\cite{wan-exp}
The error in ${\cal O}(\R)$, and therefore in the interpolation,
should accordingly also decrease exponentially beyond some supercell size.

We report numerical tests for two 
cases: the isolated set of four valence bands in Pb, and
the low-lying bands of Li, using the same WFs as in Secs.~\ref{sec:examples}
and~\ref{sec:hall}, respectively.
The band energies are computed via both 
Wannier interpolation and non-self-consistent diagonalization of
the planewave Hamiltonian on a $200\times 200 \times 200$ BZ grid.
For Li we collect data from the bottom of the inner energy window to
0.5eV below the top of the inner energy window; points close to the top
of the inner window may show larger discrepancies, as they result
from an interpolation between $\q$-points inside and outside the inner window.
Fig.~\ref{fig:band-conv} shows several measures of the difference in the 
energies as a function of $N_0^{1/3}$. In both cases we find that the
error decreases exponentially for $N_0\gtrsim 10$.
It is particularly reassuring that this occurs in Li, since the decay 
properties of disentangled WFs has yet to be
investigated thoroughly, and they probably fall outside
the scope of existing formal proofs of exponential decay.

Finally, we examine the optimal choice of supercell in which to define
the set of lattice vectors $\R$ for the Fourier transform in
Eq.~(\ref{eq:op_k}). 
To illustrate the discussion introduced earlier in the vicinity
of Eq.~(\ref{eq:op_k}), we compare the
results for parallelepipedal and Wigner-Seitz supercells. In the upper part of
Fig.~\ref{fig:wigner_vs_par} we compare the interpolated energy
bands for a
$4\times 4 \times 4$ grid of $\q$-points. For such a sparse $\q$-grid the
interpolated bands do not agree precisely with the exact {\it ab-initio} bands 
from a non-self-consistent diagonalization of the planewave Hamiltonian; this 
is most noticeable in the deviation of the curvature of the three upper bands
between K and  $\Gamma$. However, it is clear that the 
Wigner-Seitz supercell yields significantly better results than the
parallelepipedal cell. This is most clear for the upper band from L to
$\Gamma$, which displays large oscillations for the parallelepipedal
cell.  In the lower portion of Fig.~\ref{fig:wigner_vs_par}, we show 
several measures of the error in the interpolated bands as a function
of $N_0^{1/3}$; for any
given $\q$-grid the Wigner-Seitz cell gives the more accurate
results. The superiority of the  Wigner-Seitz choice
can be easily understood, as
it ensures the largest minimum distance between a WF and its periodic images.

\end{document}